\documentclass[11pt]{article}
\usepackage{graphicx}
\usepackage{amsmath}
\usepackage{etoolbox}
\usepackage[margin=0.8in]{geometry}
\usepackage{amsfonts, amssymb}
\usepackage{float}
\usepackage{hyperref}
\usepackage{cleveref}
\usepackage{caption}
\usepackage{pdfpages}
\usepackage{multicol}

\usepackage{cite}
\makeatletter
\newcommand{\ro}{\vec{r}\@ifnextchar{^}{\,}{}}
\makeatother
\patchcmd{\thebibliography}{\chapter*}{\section*}{}{}
\begin{document}

	\begin{center}\textbf{{\Large Direct calculation of time varying Aharonov Bohm effect \vspace{0.3cm}\\
		}}
	 S.Rai Choudhury$^1$ \& Shobhit Mahajan$^{*2}$ \vspace{0.3cm}\\
	 
	 	{\small $^*$Email id: shobhit.mahajan$@$gmail.com}\\
	 	
		{\small $^1$ Indian Institute of Science Education and Research, Bhopal 462066, India\\
			$^2$ Department of Physics \& Astrophysics, University of Delhi, Delhi 110007, India\\
		}
	\end{center}
\begin{abstract}
	The Aharonov-Bohm effect (ABE) for steady magnetic fields is a well known phenomenon. However, if the current in the infinite solenoid that creates the magnetic field is time-dependent, that is in the presence of both magnetic and electric fields, there is no agreement whether the effect would be present. In this note, we try to investigate time varying ABE by a direct calculation in a set-up with a weak time dependent magnetic field. We find that the electric field arising out of the time-varying magnetic field in the path of the electrons does not enter the action integral but only changes the path of the electron from the source to the slits and then on to the detector. We find a frequency dependent AB phase shift. At low frequencies the result smoothly approaches the one for a constant field as the frequency tends towards zero. On the other hand, for high frequencies such that the AB-phase induced in the path of the wave packet oscillates rapidly, the net effect will be very small which is borne out by our results. \\
\end{abstract}
\section{Introduction}
The Aharonov-Bohm effect (henceforth ABE) for steady magnetic fields is well established both theoretically and experimentally \cite{aharanov,tonomura, osakabe}. The ABE typically involves changes in a two slit interference pattern of charged particles, usually electrons, that go through  either side of a solenoid (ideally infinite) carrying a current. If the current in the infinite solenoid that creates the magnetic field is time independent, then the electrons move in a field free region. The effect then is due solely to the vector potential arising due to the solenoid current and this is referred to in literature as type I Aharonov-Bohm effect (ABE I). If however the solenoid current is time dependent then there are finite electric and magnetic fields in the path of the electrons and the change in the interference pattern is caused both by the vector potential and the associated fields. This effect is called the type II Aharonov-Bohm effect (ABE II) in literature and the distinction between the two ABE is discussed by Batelaan and Tonomura  \cite{batelaan}.
There is no agreement in the literature whether ABE II will  result in  a non-zero shift in the interference pattern or that the net effect due to the electric field, magnetic field and the vector potential will result in no shift of the interference pattern. The state of affairs is well summarised in a recent paper \cite{jian}.\\

The essential point of disagreement is whether the interaction of the electron with the electric field created by the time-dependent magnetic field cancels the magnetic ABE leading to no fringe shift. This has been claimed by Van-Kampen \cite{vankampen} and more recently by Singleton and Vagenas \cite{singleton}. This however is not supported by the work of Roy and Singh \cite{roy} who show that the dynamics is determined by the time changing flux and an additive constant parameter and so the effect due to the time-dependent flux does not automatically have cancellation between the electric and magnetic ABE. On the other hand, Brown and Home \cite{brown} claim that there will be partial cancellation between the electric and magnetic ABE with the magnetic effect being bigger. Jing et. al.\cite{jian} and Lee et. al. \cite{lee} have presented different arguments to show that a non-trivial fringe shift would be there in a time-dependent ABE contrary to claims by others \cite{vankampen},\cite{singleton}.\\
Experimentally, the situation is equally controversial. A very old experiment \cite{lmarton}, where the existence of a time varying magnetic field was realised, observed no fringe shift which was interpreted as confirmation of the theoretical result of \cite{singleton} as analysed in \cite{jmarton}. Another experiment \cite{yuv} also reported a null result but it is claimed in \cite{jmarton} that there are problems associated with this experiment. \\
The purpose of this note is to understand the time varying ABE by a direct calculation. The ABE is a quantum effect and we shall follow the simplest quantum mechanical apparatus, namely the path integral method to directly calculate the path difference in an ABE set-up with a weak time-dependent magnetic field. \\

\section{AB effect with time dependent $\vec B$}
\begin{figure}
	\begin{center}
		\includegraphics[height=10cm,keepaspectratio,scale=0.5]{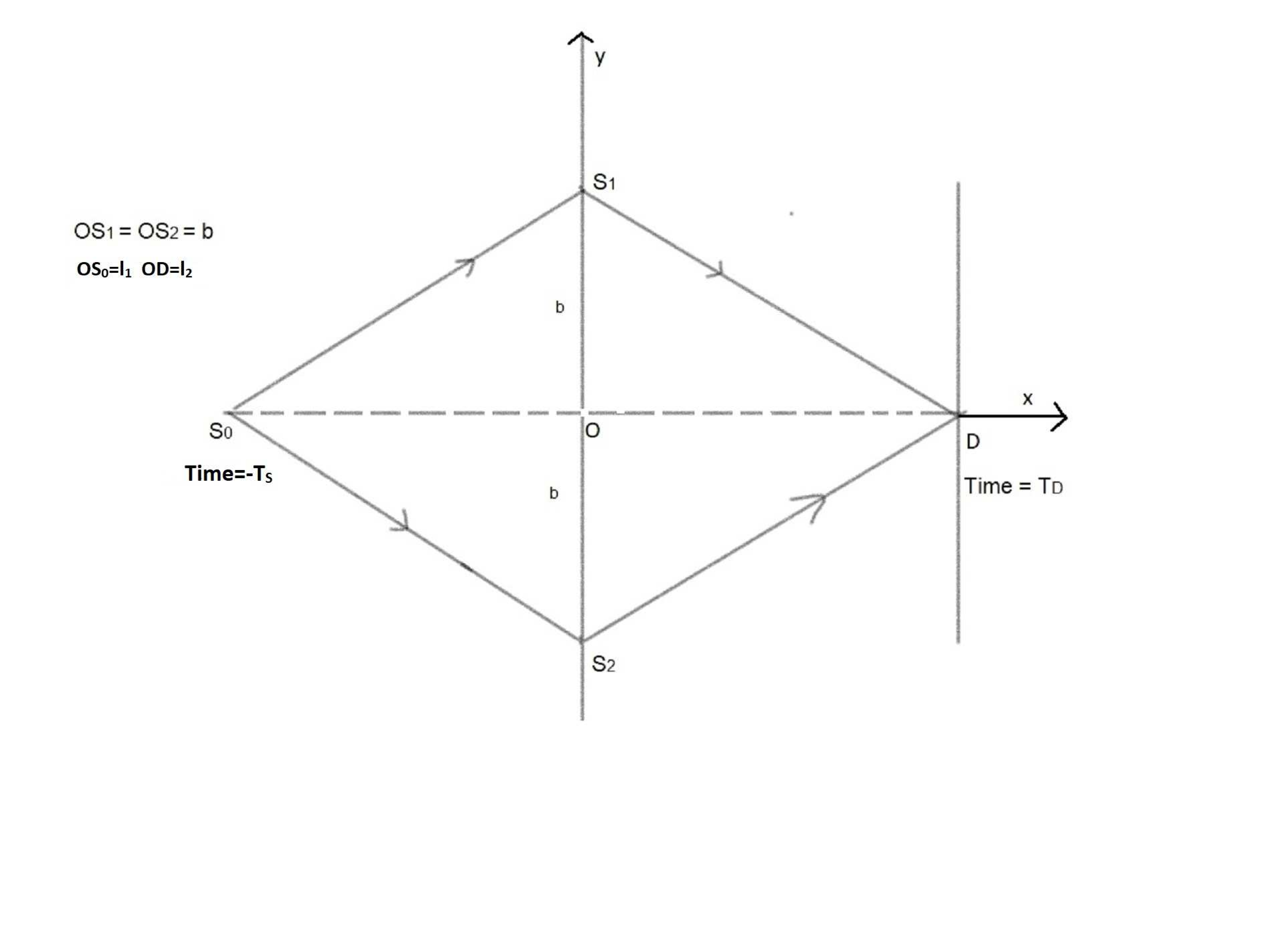}
		\captionof{figure}{Schematic Diagram of the AB effect set-up}
		\label{fig1}
	\end{center}
\end{figure}

Our main interest is to see what happens in an AB-effect scenario if the magnetic field varies in a sinusoidal fashion rather than being a constant in time. Consider an AB set-up as shown in Fig \ref{fig1}.\\

Point-like slits $S_1$ and $S_2$ are situated symmetrically (at a distance $b$ from the origin)on the $y$-axis passing through the origin $O$. Barring the slits, the barrier along the $y$-axis, passing through $O$ is impenetrable. $D$ is the point of observation on the screen. The line $SOD$ from the source $S$ to the observation point $D$ is along the $x$-axis with $y=0$. \\

In the absence of any fields, we have a two slit interference  with no phase shift between the two waves along $S_0S_1D$ and $S_0S_2D$. We now consider an infinite solenoid along the $z$ axis passing through $O$ carrying a time-dependent current. We are interested in calculating the amplitude at $D$ at time $T_D$,when an electron is emitted at $S_0$ at time $-T_S ( T_S>0)$. \\

The $z$ coordinate plays no role and we assume that $S_0, S_1, S_2$ and the impenetrable barrier all extend along the $z$ axis, so that Fig \ref{fig1} shows a section of the set-up at $z=0$.\\

We consider a surface current density 

\begin{equation}\label{eq1}
I_S(t)= \lambda I_0 \sin (\omega t + \frac{\pi}{2})
\end{equation}

the factor $\lambda$ is inserted to keep track of the order of the field created and $\delta=\frac{\pi}{2}$ such that $\omega=0$ corresponds to the steady current case. The flux in the steady current case, which is restricted to the inside of the solenoid,  is denoted by 

\begin{equation}\label{eq2}
\lambda \Phi_s= \lambda \mu_0 I_0 \pi R^2
\end{equation}

where $R$ is the radius of the solenoid. The vector potential due to such a current in the Lorentz gauge is given by \cite{abbott} 

$$ \vec{A}(\vec{r},t)= A(\vec{r},t) \hat{\phi}$$

\begin{equation}\label{eq3}
A(r,t)= - \frac{\lambda \Phi_s}{2R} J_1\left(\frac{\omega R}{c}\right) \left[ \sin(\omega t+ \frac{\pi}{2})N_1\left(\frac{\omega r}{c}\right)+ \cos(\omega t + \frac{\pi}{2}) J_1\left(\frac{\omega r}{c}\right)\right]
\end{equation}

We shall restrict ourselves to ordinary alternating currents. We shall also assume that $r$ is of order a few centimeters, which is typical of most experimental setups. Under these conditions, $\frac{\omega r}{c} \ll 1$ and we can approximate the vector potential 

\begin{eqnarray}\label{eq4}
A(r,t) &=& - \frac{\lambda \Phi_s}{2R} \left(\frac{\omega R}{2c}\right) \left[ \cos(\omega t)\left(- \frac{2 c}{\pi \omega r}\right) - \sin(\omega t) \frac{\omega r}{c}\right] \nonumber \\
&\approx& \frac{\lambda \Phi_s}{2 \pi} \frac{1}{r} \cos(\omega t)
\end{eqnarray}

neglecting $\frac{1}{c^2}$ terms. \\

The  complete  Lagrangian of an electron in an electromagnetic field is given by
$$ L = \frac{m}{2} \dot{\vec r}^{{\,}2} + q \vec A(\vec r(t),t)\cdot \dot{\vec r}(t) - q \phi(\vec r, t)$$


where $\vec A$ and $\phi$ respectively are the vector and scalar potentials. We choose for the purpose of this present calculation to work in the gauge $\phi=0$, so that the vector potential is as given in Equation \ref{eq3}. Later, in the paper we will comment on this choice of gauge. The relevant path integral is then given by


\begin{equation}\label{eq5}
K(\vec r_D,T_D;\vec r_{S_0},-T_S)= \mathcal{N} \int\limits_{S_0}^{D} \mathcal{D}\vec r(t) \exp \left[ \frac{i}{\hbar} \int\limits_{-T_S}^{T_D} L(\vec r(t), \dot{\vec r}(t)) dt \right]
\end{equation}

where the integral over $\vec{r(t)}$ is over all paths with $\vec{r}(-T_S)=\vec{r}(S_0)$ and $\vec{r}(T_D)=\vec{r}_D$. $\mathcal{N}$ is a normalization constant. The barrier containing the slits is impenetrable except at the slits $S_1$ and $S_2$. Thus, the paths divide themselves into two categories- ones which pass above $O$ and pass through $S_1$ and ones passing below $O$ and passing through $S_2$. The action integral thus is 

\begin{equation}\label{eq6}
S= \int\limits_{-T_S}^{T_D} L(\vec{r}(t),\dot{\vec r}(t), t) dt
\end{equation}

We can write the path integral as 

\begin{eqnarray}\label{eq7}
K(\vec r_D,T_D;\vec r_{S_0},-T_S)&=& \mathcal{N} \int\limits_{S_0}^{D} \mathcal{D}\vec r(t) \exp\left[\frac{i}{\hbar}S\right] \nonumber \\
&=& \mathcal{N} \int\limits_{S_0 \rightarrow S_1 \rightarrow D} \mathcal{D}\vec r(t)  \exp\left[\frac{i}{\hbar}S_U\right] + \mathcal{N} \int\limits_{S_0 \rightarrow S_2 \rightarrow D} \mathcal{D}\vec r(t)  \exp\left[\frac{i}{\hbar}S_L\right] 
\end{eqnarray}

where $S_U$ and $S_L$ refer to the action along the paths through $S_1$ and $S_2$ respectively. \\

Considering $S_U$, we can write

\begin{equation}\label{eq8}
\vec{r_{U}}(t) = \vec{r_{U}}^{cl}(t) + \vec{y_U}(t)
\end{equation}

where $\vec{r_{U}}^{cl}(t)$ is the classical trajectory from $S_0$ to $D$ going through $S_1$ and $\vec{y_U}(t)$ are deviations from it. As usual, we retain terms upto quadratic in $\vec{y}(t)$ and $\dot{\vec y}(t)$ while expanding $S_U$. Doing this, we get 

\begin{equation}\label{eq9}
\exp\left[\frac{i}{\hbar}S_U\right] = exp\left[\frac{i}{\hbar}S^{cl}_U\right] \exp\left[\frac{i}{\hbar} \int\limits_{-T_S}^{T_D} dt \left( \frac{m}{2}\dot{\vec{y}}_U^{{\,}2}(t) + q \frac{\partial A_i}{\partial r_j}y_j(t)\dot{y_i}(t) + q \frac{\partial^2 A_i}{\partial r_j \partial r_k} y_j y_k\dot{r_i}^{cl}(t)\right)\right]
\end{equation}

where all the derivatives of $\vec A$ are to be evaluated at $\vec{r}(t)=\vec{r}_{cl}(t)$ and $S_U^{cl}$ is $S$ at $\vec{r}(t)=\vec{r}_{cl}(t)$. We can write $\vec{r}_{cl}(t)$ as 

\begin{equation}\label{eq10}
\vec{r}_{cl}(t)=\vec{r}_{cl}^{\,0}(t)+ \vec{\xi}(t)
\end{equation}

where $\vec{r}_{cl}^{\,0}(t)$ is the classical trajectory for $\vec A=0$. $\vec{\xi}(t)$ is then clearly of $\mathcal{O}(\lambda)$. Then,

\begin{equation}\label{eq11}
S^{cl}_U(\vec{r}(t)) = \int\limits_{S_0S_1D} \left[\frac{m}{2}\left((\dot{\vec{r}}_{cl}^{\,0}(t)^2)+ \dot{\vec{\xi}}^{\,2}(t) + 2 \dot{\vec{r}}_{cl}^{\,0}(t) \cdot \dot{\vec{\xi}}(t)\right) + q \vec A(\vec{r}_{cl}^{\,0}(t),\dot{\vec{r}}_{cl}^{\,0}(t),t) \cdot \dot{\vec{r}}_{cl}^{\,0}(t)\right] dt
\end{equation}

neglecting $\mathcal{O}(\lambda^2)$ terms. But $\dot{\vec{r}}(t)^{\,0}$, the classical velocity for $\vec A=0$ is independent of time along  each of the straight line sectors  and  thus the second term in the rhs of Eq(\ref{eq11}) vanishes since  $\xi(t)$ vanishes at the limits of integration. Also $\vec{\xi}^2$ is of $\mathcal{O}(\lambda^2)$ which we are neglecting. Thus we get 

\begin{equation}\label{eq12}
S^{cl}_U(\vec{r}_{cl}^{\,0}(t)) \approx \int\limits_{S_0S_1D} dt \left[ \frac{m}{2}(\dot{\vec{r}}_{cl}^{\,0}(t)^2)+ q \vec A(\vec{r}_{cl}^{\,0}(t),t) \cdot \dot{\vec{r}}_{cl}^{\,0}(t)\right]
\end{equation}

Thus, with the neglecting of $\mathcal{O}(\lambda^2)$ terms, we can write, 

\begin{equation}\label{eq13}
\exp\left[\frac{i}{\hbar}S_U\right] = exp\left[\frac{i}{\hbar}S^{cl}_U(\vec{r}_{cl}^{\,0}(t), \dot{\vec{r}}_{cl}^{\,0},t)\right] \exp\left[\frac{i}{\hbar} \int\limits_{-T_S}^{T_D} dt \left( \frac{m}{2}\dot{\vec{y}}_U^{{\,}2}(t) + q \frac{\partial A_i}{\partial r_j}y_j(t)\dot{y_i}(t) + q \frac{\partial^2 A_i}{\partial r_j \partial r_k} y_j y_k\dot{r_i}^{cl}(t)\right)\right]
\end{equation}

with all the derivatives evaluated at $\vec{r}(t)=\vec{r}_{cl}^{\,0}(t)$.\\

We will shortly show that for field strength that causes fringe number shift $\sim 1$,  
the quantum fluctuation terms quadratic in $\vec{y}(t)$ involving $\vec{A}(\vec r, t)$ are negligible. Given this, we get

\begin{equation}\label{eq14}
\mathcal{N} \int\limits_{S}^{D} \mathcal{D}\vec r(t) \exp\left[\frac{i}{\hbar}S_U\right] = \mathcal{N} \exp\left[\frac{i}{\hbar} S_U^{cl}(\vec{r}_{cl}^{\,0}, \dot{\vec{r}}_{cl}^{\,0},t)\right] \int\limits_{S_0S_1D} \mathcal{D}\vec r(t)  \exp\left[\frac{i}{\hbar}\int\limits_{-T_S}^{T_D} dt \frac{m}{2} \dot{\vec{y}}_U^{\,2}\right]
\end{equation}

Since $\dot{\vec{r}}_{cl}^{\,0}(t)$ is time independent for each straight line section of the classical path, this can be easily computed. \\


The point to note is that in the gauge we have chosen to work with, the electric field enters non trivially in our calculations. Thus, it enters first by both  changing the classical path due to the classical force arising out of the electric field. Secondly, it enters indirectly through the value of the path integral  since we evaluate the integral with explicit time dependence of the vector potential  $\vec A(\vec r, t(\vec r))$. Since the electric field $\vec E$ at every point of the path is given by $\vec E = - \frac{\partial \vec A(\vec r, t(\vec r))}{\partial t}$, the electric field is built into the path integral.\\

We will now show that the quantum effects caused by $\vec A(\vec r,t)$ are small. We shall consider typical strengths of $\vec A(\vec r,t)$ which for the static case cause fringe shifts, $\Delta n_S \sim 1$. In the steady case, the phase arising out of $\vec A(\vec r, t)$ is 

\begin{equation}\label{eq15}
i \frac{q}{\hbar} \oint \vec A \cdot \vec{dr} = \frac{i q}{\hbar} \frac{\lambda \Phi_s}{2 \pi r} 2 \pi r 
\end{equation}
so that 
\begin{equation}\label{eq16}
\frac{1}{2 \pi} \lambda \frac{q}{\hbar} \Phi_s = \Delta n_S \approx 1
\end{equation}

The scale of quantum fluctuations $\vec y$ in Eq(\ref{eq13}) is set by the kinetic term

\begin{equation}\label{eq17}
|\dot{\vec{y}}|^2  \lessapprox \frac{2 \hbar}{m T} \qquad \qquad \qquad  (T \sim T_S, T_S)
\end{equation}
and hence 
\begin{equation}\label{eq18}
|\vec{y}|  \lessapprox \sqrt{\frac{2 \hbar T}{m }} 
\end{equation}

Consider now the $\left(\frac{\partial A_i}{\partial r_j}\right)$ term in the rhs of Eq(\ref{eq13}). Its order of magnitude is 

\begin{eqnarray}\label{eq19}
\frac{q}{\hbar} \bigl|\frac{\partial A_i}{\partial r_j}\bigl| \big|y\big|\big|\dot{y}\big| &<& \frac{q}{\hbar} \bigl|\frac{\lambda \Phi_s}{2 \pi R^2}\bigl| \frac{2 \hbar}{m} \nonumber\\
&\approx& \Delta n_s \frac{2 \hbar}{m R^2}
\end{eqnarray}
where $R \sim b, l$ is a few cms. \\

The ratio of this to $\frac{m}{2} |\dot{\vec{y}}|^2$, which is of $\mathcal{O}(\frac{1}{T})$ is 

\begin{equation}\label{eq20}
\frac{\Delta n_s}{m R^2}\hbar T \approx (\Delta n_s) \frac{1}{R} \left(\frac{\hbar}{m v}\right) \approx (\Delta n_s) \frac{\lambda}{R}
\end{equation}
where now $\lambda$ is the deBroglie wavelength of the electron and $v\sim \frac{R}{T}$ is of the order of magnitude of the speed of the electron. 

For electrons of kinetic energy $\sim 10$ eV, $mv \sim (10^{-27} \times 10^{-12})^{\frac{1}{2}}$ that is $10^{-19} \text{ gm cm s}^{-1}$ and thus $\lambda \sim 10^{-8}$ cm which is $\ll R$. Similar arguments apply to the other terms involoving $A$ and $y$ in $S_U$ in Eq(\ref{eq13}). Thus the quantum fluctuation is very much dominated by the kinetic term.\\

The free particle path integral with quadratic fluctuations can be calculated easily \cite{feynman}. In our case, we get 

\begin{equation}\label{eq21}
\exp\left[\frac{i}{\hbar} S_U\right] \approx \left(\frac{m}{2 \pi i \hbar T_S}\right)^{\frac{1}{2}} \exp \left[\frac{i m (l^2+b^2)}{2 \hbar T_S}\right] \left(\frac{m}{2 \pi i \hbar T_D}\right)^{\frac{1}{2}} \exp \left[\frac{i m (l^2+b^2)}{2 \hbar T_D}\right] \exp\left[\frac{iq}{\hbar}\int\limits_{-T_S}^{T_D} \vec A(\vec{r}_{cl}^{\,0}(t),\dot{\vec{r}}_{cl}^{\,0}(t),t)\cdot \dot{\vec{r}}_{cl}(t) dt\right]
\end{equation}

We can easily see that $\dot{\vec{r}}_{cl}^{\,0}(t)$ in the $SS_1$ and $S_1D$ sections is constant. We assume that the incident wave packet reaches the slits $S_1$ and $S_2$ at the same time which we take as the zero of time. We can easily evaluate this
For $S_0S_1$:	
$$ \dot{\vec{r}}_{cl}^{\,0}(t)=\left(\frac{l_1}{T_S},\frac{b}{T_S},0\right)$$
For $S_1D$:
$$ \dot{\vec{r}}_{cl}^{\,0}(t)=\left(\frac{l_2}{T_D},-\frac{b}{T_D},0\right)$$

Using Eq(\ref{eq4}), we can evaluate the last integral in Equation(\ref{eq21}) along the path $S_0S_1D$. 

\begin{equation}\label{eq22}
I_U= \frac{i q}{\hbar} \int\limits_{-T_S}^{T_D} \vec A(\vec{r}_{cl}^{\,0}(t),\dot{\vec{r}}_{cl}^{\,0}(t),t)\cdot \dot{\vec{r}}_{cl}(t) dt
\end{equation}

Along $S_0S_1$:
\begin{equation}\label{eq23}
\dot{\vec{r}}_{cl}^{\,0}(t)=\left(\frac{l_1}{T_S},\frac{b}{T_S},0\right)
\end{equation}

\begin{equation}\label{eq24}
\vec{r}_{cl}^{\,0}(t)=\left(\frac{l_1 t}{T_S},\frac{b(t+T_S)}{T_S},0\right)
\end{equation}

Along $S_1D$:
\begin{equation}\label{eq25}
\dot{\vec{r}}_{cl}^{\,0}(t)=\left(\frac{l_2}{T_D},-\frac{b}{T_D},0\right)
\end{equation}

\begin{equation}\label{eq26}
\vec{r}_{cl}^{\,0}(t)=\left(\frac{l_2 t}{T_D},\frac{b(T_D-t)}{T_D},0\right)
\end{equation}

Throughout the path, we have 

$$ \hat{\theta} = \frac{-y \hat x + x \hat y}{r_{cl}^{\,0}}$$

Thus, along $S_0S_1$, 

\begin{eqnarray}\label{eq27}
\left(\hat{\theta}\cdot \dot{\vec{r}}_{cl}^{\,0}\right) &=& \frac{-l_1b(T_S+t)+l_1bt}{T_S r_{cl}^{\,0}} \nonumber \\
&=& \frac{-l_1b}{ r_{cl}^{\,0}} \nonumber \\
&=& \frac{-bl_1 T_S}{\left[l_1^2t^2+b^2(T_S+t)^2\right]^{1/2}} 
\end{eqnarray}

and along $S_1D$,

\begin{eqnarray}\label{eq28}
\left(\hat{\theta}\cdot \dot{\vec{r}}_{cl}^{\,0}\right) &=& \frac{-l_2b(T_D-t)-l_2bt}{T_D r_{cl}^{\,0}} \nonumber \\
&=& \frac{-l_2b}{r_{cl}^{\,0}} \nonumber \\
&=& \frac{-bl_2 T_D}{\left[l_2^2t^2+b^2(T_D-t)^2\right]^{1/2}} 
\end{eqnarray}

Putting all this in the integral above, we get 
\begin{align}\label{eq29}
I_U =& \left[\left(\frac{iq}{\hbar}\right)\left[\frac{\lambda \Phi_s}{2 \pi}\right]\right]\left[-T_S\int\limits_{-T_S}^{0} dt \frac{\cos(\omega t) b l_1}{[l_1^2t^2+b^2(T_S+t)^2]} - T_D \int\limits_{0}^{T_D} dt \frac{\cos(\omega t) b l_2}{[l_2^2t^2+b^2(T_D-t)^2]}\right] \nonumber \\
=&\left[\left(\frac{iq}{\hbar}\right)\left[\frac{\lambda \Phi_s}{2 \pi}\right]\right]\left[-T_S\int\limits_{-T_S}^{0} dt \frac{\cos(\omega t) b l_1}{t^2(l_1^2+b^2)+2 t b^2T_S + b^2T_S^2} - T_D \int\limits_{0}^{T_D} dt \frac{\cos(\omega t) b l_2}{t^2(l_2^2+b^2)-2 t b^2T_D +b^2T_D^2}\right]\nonumber \\
=& \left[\left(\frac{iq}{\hbar}\right)\left[\frac{\lambda \Phi_s}{2 \pi}\right]\right]\left[-T_S\int\limits_{-T_S}^{0}\left(\frac{bl_1}{b^2+l_1^2}\right) \frac{dt \cos(\omega t)}{\left[t+\frac{b^2T_s^2}{b^2+l_1^2}\right]^2 + \left[\frac{b^2T_S^2}{b^2+l_1^2} -\frac{b^4T_S^2}{(b^2+l_1^2)^2}\right]}\right] \nonumber \\
& - \left[\left(\frac{iq}{\hbar}\right)\left[\frac{\lambda \Phi_s}{2 \pi}\right]\right]\left[ T_D\int\limits_{0}^{T_D}\left(\frac{bl_2}{b^2+l_2^2}\right) \frac{dt \cos(\omega t)}{\left[t-\frac{b^2T_D^2}{b^2+l_2^2}\right]^2 + \left[\frac{b^2T_D^2}{b^2+l_2^2} -\frac{b^4T_D^2}{(b^2+l_2^2)^2}\right]}\right] \nonumber \\
=&  \left[\left(\frac{iq}{\hbar}\right)\left[\frac{\lambda \Phi_s}{2 \pi}\right]\right]\left[-T_S\left(\frac{bl_1}{(b^2+l_1^2)}\right)\int\limits_{-T_S+\frac{b^2T_S}{b^2+l_1^2}}^{\frac{b^2T_S}{b^2+l_1^2}}\frac{dy \cos\left[\omega(y-\frac{b^2T_S}{(b^2+l_1^2)^2})\right]}{y^2+\frac{b^2l_1^2T_S^2}{(b^2+l_1^2)^2}}\right] \nonumber \\
& - \left[\left(\frac{iq}{\hbar}\right)\left[\frac{\lambda \Phi_s}{2 \pi}\right]\right]\left[-T_D\left(\frac{bl_2}{(b^2+l_2^2)}\right)\int\limits_{\frac{-b^2T_D}{b^2+l_2^2}}^{\frac{l_2^2T_D}{b^2+l_2^2}} \frac{dy\cos\left[\omega(y+\frac{b^2T_D}{(b^2+l_2^2)^2})\right]}{y^2+\frac{b^2l_2^2T_D^2}{(b^2+l_2^2)^2}}\right] 
\end{align}

$I_L$  is defined as in Eq(\ref{eq22})  for  $I_R$  with $\vec{r}_{cl}^{\,0}$ along the lower path $S_0 S_2 D$.  $I_L$ can be obtained from $I_U$  with the substitution  $b \rightarrow -b$ and hence  $I_L = -I_U$
To get the steady result, $\omega =0$, we can use Eq(\ref{eq29}).

\begin{align}\label{eq30}
I_U=&\left(\frac{iq\lambda \Phi_s}{2 \pi \hbar}\right)\left[\left[\frac{bl_1}{b^2+l_1^2}\right]\left[-T_S\left(\frac{(b^2+l_1^2)^2}{b^2l_1^2T_S^2}\right)^{1/2}\arctan\left(\frac{y(b^2+l_1^2)}{bl_1T_S}\right)\Biggl|_{\frac{-T_Sl^2}{(b^2+l_1^2)}}^{\frac{T_Sb^2}{(b^2+l_1^2)}}\right] \right] \nonumber \\
&- \left(\frac{iq\lambda \Phi_s}{2 \pi \hbar}\right)\left[\left[\frac{bl_2}{b^2+l_2^2}\right]\left[T_D\left(\frac{(b^2+l_2^2)^2}{b^2l_2^2T_D^2}\right)^{1/2}\arctan\left(\frac{y(b^2+l_2^2)}{bl_2T_D}\right)\Biggl|_{\frac{-T_Db^2}{(b^2+l_2^2)}}^{\frac{T_Dl^2}{(b^2+l_2^2)}}\right] \right] \nonumber \\
=& - \left[\frac{i q \lambda \Phi_s}{2 \pi \hbar}\right]\left[\left(\arctan\left(\frac{b}{l_1}\right)- \arctan\left(\frac{-l_1}{b}\right)\right)+\left(\arctan\left(\frac{l_2}{b}\right)\right)- \arctan\left(\frac{b}{l_2}\right)\right] \nonumber \\
=& - \left[\frac{i q \lambda \Phi_s}{2 \pi \hbar}\right]\left[\frac{\pi}{2}+\frac{\pi}{2}\right] \nonumber \\
=& - \left[\frac{i q \lambda \Phi_s}{2 \hbar}\right]
\end{align}

The analog of Eq(\ref{eq21}) for the lower path will have the same factors in front of the last exponential phase factor. This last exponential factor for the upper and lower paths thus will be  $\exp (I_U)$  and $\exp(I_L)$ respectively. From Eq(\ref{eq7}),  the interference pattern will be dictated by the factor  $[1 + \exp(I_U-I_L)]$.  The central point of the detector $D$ that we are considering  will, for the case  $\vec A =0$, have zero phase difference between the upper and lower paths and hence be a bright spot for the interference pattern which can be called fringe number $n = 0$. The other points on the detector can be characterized by a `fringe number’ $n$ such that  the phase difference there is $2\pi n$.\\

In the presence of  a non vanishing $\vec A$ , the central point $D$ will no longer correspond to the $n=0$ fringe but will be shifted as the fringe number  $\Delta  n_S \equiv \frac{I_U-I_L}{2\pi i }$.\\

We see that for $\omega=0$, 

\begin{equation} \label{eq31}
\frac{1}{2 \pi i} (I_L-I_U) = \frac{q}{2 \pi \hbar} \lambda \Phi_s=\Delta n_S,
\end{equation}

where  the subscript $S$ refers to a steady  value of $\vec A$.\\

For   $ \omega  \neq  0 $ the  integrals have to be evaluated numerically. The only place where  $\omega  $  enters in Eq(\ref{eq29}) is in the cosine factors in integrands  in the rhs.  Hence for $ \omega \neq 0$, the fringe number shift  $(\Delta n)_{\omega} $ is


\begin{align}\label{eq32}
(\Delta n)_{\omega}=& (\Delta n)_S \left(\frac{1}{\pi}\right)\left[-T_S\left(\frac{bl_1}{(b^2+l_1^2)}\right)\int\limits_{-T_S+\frac{b^2T_S}{b^2+l_1^2}}^{\frac{b^2T_S}{b^2+l_1^2}}\frac{dy \cos\left[\omega(y-\frac{b^2T_S}{(b^2+l_1^2)^2})\right]}{y^2+\frac{b^2l_1^2T_S^2}{(b^2+l_1^2)^2}}\right] \nonumber \\
& - (\Delta n)_S \left(\frac{1}{\pi}\right)\left[-T_D\left(\frac{bl_2}{(b^2+l_2^2)}\right)\int\limits_{\frac{-b^2T_D}{b^2+l_2^2}}^{\frac{l_2^2T_D}{b^2+l_2^2}} \frac{dy\cos\left[\omega(y+\frac{b^2T_D}{(b^2+l_2^2)^2})\right]}{y^2+\frac{b^2l_2^2T_D^2}{(b^2+l_2^2)^2}}\right] 
\end{align}

These integrals can be numerically evaluated for any given geometry, time scales and values of $\omega$ relevant to the experiment. However, to see the nature of $(\Delta n)_{\omega}$, we put $l_1=l_2=b=L$ and $T_S=T_D=T$. With these simplifications, Eq(\ref{eq32}) reduces to 

\begin{align}\label{eq33}
(\Delta n)_{\omega} =& (\Delta n)_S \frac{4}{\pi} \int\limits_{0}^1 dy \frac{\cos(\frac{\omega T y}{2}) \cos(\frac{\omega T}{2})}{1+y^2} \nonumber \\
=&(\Delta n)_S f(\omega T)
\end{align}

\begin{figure}
	\begin{center}
		\include{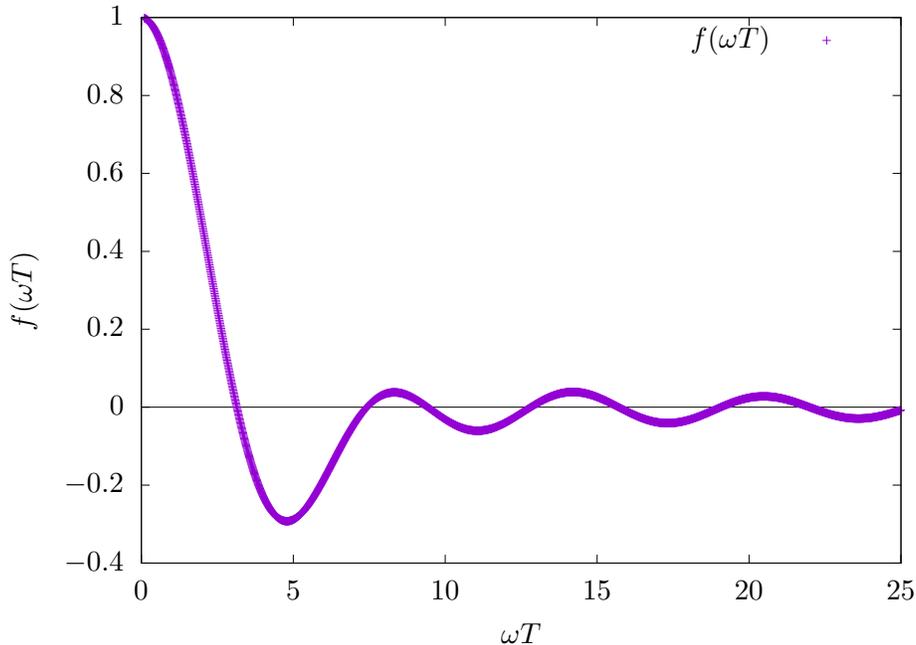}
		\captionof{figure}{The function  $f(\omega T)$   plotted against  $(\omega T)$.}
		\label{fig2}
	\end{center}
\end{figure}

The  function  $f(\omega T)$ is plotted in Figure \ref{fig2} showing non-trivial AB phase shift in the presence of a time varying solenoidal field.\\

\section{Discussion \& Conclusions}
Summarizing, we have directly, but approximately calculated the Aharanov- Bohm phase in a typical two slit electron interference pattern following the path integral method for a sinusoidal current in the solenoid. The electric field arising out of the time-varying magnetic field in the path of the electrons does not enter the action integral but only changes the path of the electron from the source to the slits and then on to the detector. Our results show that there is a frequency dependent AB- phase shift. At low frequencies the result smoothly approaches the one for constant field as the frequency tends towards zero. This is to be expected since if during the time of transit of the wave function packet from the source to the detector, there is no appreciable change in the magnetic field, the result effectively will be the same as for constant fields. On the other hand, for high frequencies such that the AB-phase induced in the path of the wave packet  oscillates  rapidly, the net effect will be very small which is borne out by our results.\\

A comment on our choice of gauge is in order. We have used the results of the potentials and fields evaluated in the gauge with  $\phi=0$ by Abbott and Griffiths  \cite{abbott}. Making a gauge transformation and then performing a numerical evaluation becomes numerically very cumbersome and we have not been able to it. The question of gauge invariance for ABE II has been investigated by examining the Stokes’ theorem relevant to ABE II by Macdougall and Singleton \cite{macdougall}  and by Marcovitch et. al. \cite{marcovitch}. These investigations  reveal that both the electric and magnetic fields get involved in working out the final result of the Stokes’ integral. In our approach both the fields play a role in the evaluation of the path integral. In particular for time varying currents in the solenoid, the electric field affects the classical path as well as the value of the path integral since the vector potential at each point of the path involved is time dependent.\\

We would also comment on our  calculation and result in relation to two other relevant investigations of ABE II. Bright \textit{et al.}\cite{bright} have investigated time dependent ABE II wherein the time dependence of the fields comes from an  electromagnetic wave background. The cancellation of the electric and magnetic fringe shifts was shown to be approximate. Although the setup is different from the case we have considered (Fig \ref{fig1}), our results seem to be in accord with theirs in that there is cancellation for large values of $\omega T$ but not a total one. The second aspect of ABE II is regarding the multiple connectedness of space in an ideal setup with infinite solenoid investigated by Gaveau \textit{et al}. \cite{gaveau}. The total path integral is shown to be a sum of contributions from classes of paths belonging to different winding numbers(Equations 8 \& 10 of the paper). Similar results have also been reported by Bernido and Inomata \cite{bernido} where it is shown that the leading contribution comes from the usual loop which encircles the solenoid only once. This is the one we have calculated and in view of the fact that our main interest was to show that a non trivial shift in fringes result in the setup shown Fig \ref{fig1}, it seems unlikely that the essence of our result will change by including other paths. A complete numerical calculation including all paths is a formidable numerical task which is beyond the scope of this calculation.\\

A final comment on the observability  of the AB-effect for time varying magnetic fields. The wave function collapse on the screen is statistical in  nature. If the incident flux of the electrons is almost continuous then of course the wave function modulus squared on the screen will be directly related to the pattern observed   irrespective of whether the magnetic field is time varying or time independent. If however, electrons arrive at the source point effectively one at a time as in the remarkable experiment of Tonomura \cite{tonomura}, then the final observable effect will be dependent on the time elapsed between successive electrons arriving at the source point. If the magnetic field is constant, then the phase change for each one of them is the same. On the other hand, if the time between successive electrons is much smaller than time period of the magnetic field oscillations, then the phase change for the individual wave packets will be different. On the screen , therefore there will be broadening or even total blurring of the pattern.\\

\section*{Acknowledgment}
The authors would like to thank the referees for very valuable comments. One of us (SRC) would like to thank Professor S.M.Roy for some valuable discussion.

\end{document}